# The challenge of changing deeply-held student beliefs about the relativity of simultaneity

Rachel E. Scherr[a)], Peter S. Shaffer, and Stamatis Vokos
Department of Physics, University of Washington, Seattle, WA

**Abstract**
Previous research indicates that after standard instruction students at all academic levels often construct a conceptual framework in which the ideas of absolute simultaneity and the relativity of simultaneity co-exist. This article describes the development and assessment of instructional materials intended to improve student understanding of the concept of time in special relativity, the relativity of simultaneity, and the role of observers in inertial reference frames. Results from pretests and post-tests are presented to demonstrate the effect of the curriculum in helping students deepen their understanding of these topics. Excerpts from taped interviews and classroom interactions help illustrate the intense cognitive conflict that students encounter as they are led to confront the incompatibility of their deeply-held beliefs about simultaneity with the results of special relativity.

## I. INTRODUCTION

The Physics Education Group at the University of Washington is conducting an ongoing study of student understanding of basic ideas in special relativity.[1,2] A previous article described a detailed investigation into student conceptions of time, reference frames, and simultaneity after traditional instruction.[1] We found that students often leave a standard introductory course or an advanced undergraduate course on relativity with some fundamentally incorrect beliefs about the definition of the time of an event and the construction of a reference frame.[3,4] Many seem to believe that the time of a distant event is the time at which a signal from the event is received by an observer. Thus, they treat the time ordering of two events as dependent on the location of an observer. Yet, many of these same students also have a deeply-held underlying belief that simultaneity is absolute and that when signal travel time is accounted for, all observers (in all reference frames) agree on the time order of any two events.[5] Many students thus fail to recognize one of the profound implications of special relativity for our understanding of the nature of time.

In this paper, we report on the development and assessment of curriculum designed to help students construct a meaningful understanding of the relativity of simultaneity. The initial development was guided by earlier research.[1-4] Use of the materials in the classroom revealed ways of student thinking that we had not encountered previously. These insights led to modifications that increased the effectiveness of the instruction. The current versions are the product of an iterative process, part of which is described.

Two previous articles describe conceptual change in the larger context of special relativity.[7] Those articles outline the general circumstances under which conceptual change is likely to occur, and suggest broad instructional strategies to encourage such change. This paper focuses on the effect on student learning of a particular instructional intervention and illustrates some aspects of the conceptual conflict that occurs.

## II. CONTEXT FOR RESEARCH AND CURRICULUM DEVELOPMENT

The development and testing of the instructional materials on special relativity have primarily been conducted at the University of Washington (UW). The populations have included students in the introductory calculus-based honors course (for physics majors and others with strong science and mathematics background) and students in advanced undergraduate courses (*e.g.,* the junior-level course on electricity and magnetism and a course on relativity and gravitation). All together, this study has involved the classes of six instructors. About 350 students from 12 sections of various courses have participated.

The setting for most of the work described in this article has been an extension of the tutorial system in the introductory calculus-based course. The core of the system is provided by a set of tutorials collectively entitled *Tutorials in Introductory Physics.*[8] These are designed to supplement the lectures and textbook of a traditional lecture-based course. The emphasis is on constructing concepts, developing reasoning skills, and relating the formalism of physics to the real world, not on transmitting information or solving end-of-chapter problems. The tutorials are described in other articles by our group.[9] A few key elements are described below.

Each tutorial sequence begins with a short pretest that is designed to elicit student ideas. The pretests consist of qualitative questions that require explanations of reasoning. They are typically administered after relevant lecture and textbook instruction. During the subsequent tutorial session, students work collaboratively in small groups on tutorial worksheets. These consist of a series of carefully sequenced questions intended to guide students through the reasoning necessary to develop and apply a given concept. Tutorial homework helps students apply, extend, and generalize what they have learned. Post-testing on course examinations is a crucial part of the tutorial sequence. Comparisons of student performance on the pretests and post-tests provide assessment of student learning and guide modifications to the curriculum.

## III. OVERVIEW OF THE INSTRUCTIONAL APPROACH

An understanding of the relativity of simultaneity is inextricably linked to the concept of reference frame and the operational definition of the time of a distant event. In our investigation we have observed that students often fail to interpret properly the "time of an event" and the notion of "reference frame." Many thus do not come to an understanding of these basic ideas, let alone the classic paradoxes that are typically used in instruction in special relativity. Therefore, we focus tutorial instruction on helping students develop the requisite concepts and apply the reasoning required for resolving one of the standard paradoxes: the 'train paradox.'[10]

In this article, we describe a set of two tutorials, entitled *Events and reference frames* and *Simultaneity.* The first is in the context of a single reference frame. Students are guided to develop the basic procedures that allow an observer to measure the time of a single distant event. This forms the basis for defining a reference frame as a system of intelligent observers. The tutorial then helps students extend the intuitive notion of whether or not two *local* events are simultaneous by having them develop a definition of simultaneity for events that have a spatial separation.[11,12] In the second tutorial, students examine the consequence of the invariance of the speed of light through analysis of the train paradox. They are led to recognize that resolution of the paradox requires the relativity of simultaneity as a means of preserving causality. This tutorial reinforces the equivalence of observers in a given frame in determining the time order of events. The tutorial concludes by helping students apply the relativity of simultaneity to other contexts. Students take about two hours to work through the pair of tutorials.

The tutorials are not intended as a stand-alone curriculum. The assumption is that students have been introduced to certain basic ideas (*e.g.,* the invariance of the speed of light, events, and synchronization of clocks) in other parts of the course. The content of the tutorials does not differ significantly from what is typically taught in a course on special relativity. The approach taken, however, is to help students go through the reasoning required to develop a functional understanding of the relativity of simultaneity.

The tutorials described in this paper use a variety of instructional strategies. One of these can be loosely characterized as a series of steps: *elicit, confront,* and *resolve*.[13] First, students are presented with a situation that exposes a tendency to make a particular error. Confrontation occurs when students recognize (or are led to recognize) a discrepancy between their ideas and the actual behavior of a physical system. Students are then guided through the reasoning necessary to resolve any inconsistencies.

In the discussion below, we illustrate how the tutorials attempt to address specific student difficulties. Section IV focuses on instruction to help students develop appropriate definitions for time and reference frame. Section V describes exercises to help students overcome their belief in absolute simultaneity. This section also documents how, in the process of designing curriculum, we identified some conceptual difficulties with causality and how instruction was modified to address them. The assessment of effectiveness in Section VI reports results from pretests and from post-tests administered after all tutorial instruction.

## IV. LAYING THE GROUNDWORK FOR ADDRESSING STUDENT DIFFICULTIES WITH REFERENCE FRAMES

In the previous article, we illustrated that student difficulties with the relativity of simultaneity can often be traced to beliefs about measurements of time and the meaning of reference frames.[1] We found that students at all levels tend to treat the time of an event as the time at which a signal from the event is received by an observer. Thus, they consider a reference frame as being location dependent.[5] The persistence of these beliefs about time and reference frames suggests a need for instruction that provides students with a strong foundation upon which they can draw in their study of special relativity. This is the approach taken in the tutorial, *Events and reference frames,* which focuses on time, reference frames, and simultaneity in Galilean relativity.

### A. Guiding students in the determination of the time of an event

The *Events and reference frames* tutorial begins by guiding students to formulate appropriate procedures for the measurement of the time of an event. In the first exercise, an observer wishes to know the time at which a beeper beeps but is constrained to a location far from the beeper. The observer is equipped with accurate meter sticks, and synchronized clocks, and has assistants who can help. The tutorial asks students to describe two procedures by which the observer can determine the time at which the beeper beeps: (i) using knowledge of the speed of sound in air and (ii) without knowing or measuring the speed of sound first. In this way students articulate for themselves two operational definitions for the time of a distant event: (i) an observer may record the time of arrival of the sound from an object, measure the distance to the object, and correct for the signal travel time, or (ii) an observer may place an assistant at the object and have the assistant mark the time at which it makes a sound. The exercise builds on student understanding of the finite nature of signal travel time which, as we observed during the investigation discussed in the previous paper, generally appears to be good.

### B. Guiding students in the construction of a reference frame

In a subsequent exercise, students generalize their measurement procedure for the time of an event. They are asked to devise an arrangement of observers and equipment for recording the position and time of an arbitrary event. The term *reference frame* is introduced to describe the system of observers. The term "intelligent observer" is defined as an observer who takes into account signal travel time.

### C. Guiding students in the definition of simultaneity of two events for a given reference frame

After students have constructed the concept of a "reference frame," they are asked to apply it. The context is one that we have found can elicit the belief that the time order of events "in an observer's reference frame" is the order in which signals from the events are received by the observer. Students are told that a horn is placed between an observer and a distant beeper. The observer hears a honk and a beep at the same instant. Students are asked two questions. The first is to describe a method by which the observer can measure the time separation between the emission of the two sounds in his reference frame without knowing or measuring the speed of sound first. They are also asked whether, in the observer's reference frame, the beeper beeps before, after, or at the same time as the horn honks. Students use the idea of a reference frame and the definition of the time of an event to conclude that, in order for the signals to reach the observer simultaneously, the more distant event must have occurred first. The pair of questions helps students recognize that the term "simultaneous events" does not refer to the simultaneous reception of signals generated by those events, but rather to a comparison of the time coordinates of the events as measured by a system of intelligent observers.

The ideas developed in the *Events and reference frames* tutorial seem straightforward and may appear elementary to instructors. However, evidence from post-tests suggests that this kind of instruction is necessary but not sufficient in helping students overcome their difficulties with the role of observers in a reference frame.

## V. BUILDING AN UNDERSTANDING OF THE RELATIVITY OF SIMULTANEITY

The *Events and reference frames* tutorial described above, focuses on the determination of the time of an event and the role of observers in the context of a single reference frame. In the *Simultaneity* tutorial, students draw on these ideas as they consider multiple frames.

### A. Guiding students in applying the invariance of the speed of light

After traditional instruction, most students can state that the speed of light is the same in all directions in all reference frames. We have found during instruction, however, that few students have the ability to use this knowledge to analyze relativistic scenarios.

*Single flash of light*

The *Simultaneity* tutorial begins by helping students apply the invariance of the speed of light to a simple physical situation: the isotropic propagation of the wavefront from a single flash of light as analyzed in two reference frames. Students are told that two observers, Alan and Beth, move past each other at relativistic relative speed. At the instant they pass, a spark occurs between them, emitting a flash of light. Students are shown a cross-sectional diagram for Alan's frame representing Alan, Beth, and a spherical wavefront of light a short time after the spark occurs. They are asked to identify features of the diagram that illustrate the fact that the speed of light is the same in all directions according to Alan. They are then asked to sketch a diagram corresponding to a short time later in Alan's frame. Most students recognize that a spherical wavefront shows the speed of light to be the same in all directions and sketch a larger sphere to represent the wavefront at the later time. (See Fig. 1(a) for correct diagrams.)

The students then sketch similar diagrams in Beth's reference frame. To do so, they need to recognize that Beth also observes the propagation of light to be isotropic. Thus, she is at the center of a spherical wavefront in her frame, while Alan moves relative to her. (See Fig. 1(b) for correct diagrams for Beth's frame.) This exercise is not difficult for most students. However, it lays important groundwork for the subsequent exercise.

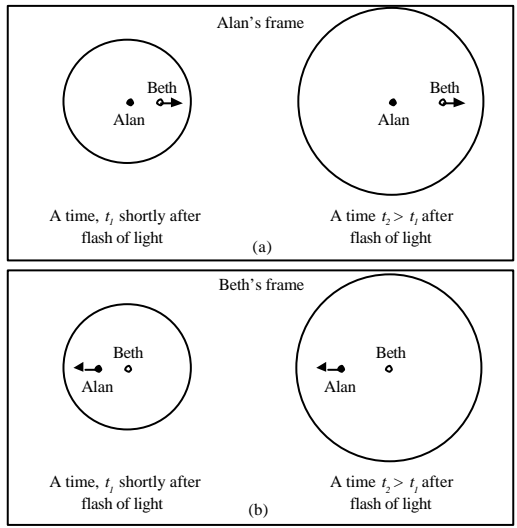

Figure 1: Diagrams from a tutorial exercise in which students apply the isotropy of free space and the invariance of the speed of light. Each circle indicates the wavefront from a brief flash of light. Students are asked to complete each diagram to show the observers and the wavefront at two different instants in each reference frame. (a) Completed diagrams for Alan's reference frame. (b) Completed diagrams for Beth's reference frame.

### *Two flashes of light (train paradox)*

In the next part of the tutorial, students begin to analyze a version of the classic train paradox, which involves two flashes of light. The paradox is summarized below.

#### *Description of train paradox*

Two flashes of lightning strike the ends of a train that is moving with uniform velocity. Both occur at the same time according to an observer at rest on the ground. In the ground frame, the observer notes that the train is moving toward the origin of one of the flashes. The observer therefore concludes that the wavefronts from the two flashes reach the center of the train at *different times*. The observer then imagines the situation in the reference frame of the train, in which the train is stationary. Knowing that the propagation of light is isotropic in all frames, the observer reasons that since the wavefronts would travel the same distance from the ends of the train to the center, they would reach the center at *the same time*. Thus, the predictions about the order in which the wavefronts reach the center of the train seem to be different in the two frames.

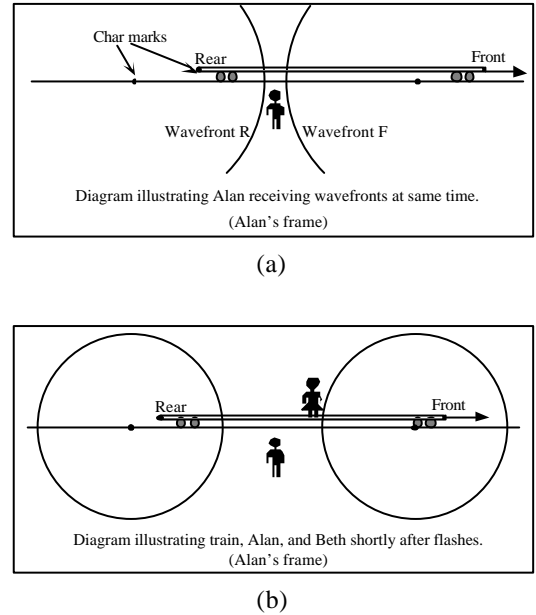

Figure 2: Diagrams of train paradox for ground-based observer. (a) Diagram given to students. The ground observer is at the center between the char marks left by two flashes of light and receives the corresponding wavefronts at the same time. (b) Example of correct diagram drawn by students to illustrate whether the front wavefront hits an observer at the center of the train before, after, or at the same instant as the rear wavefront.

The resolution of the apparent paradox is to conclude that the flashes of lightning at the ends of the train are not simultaneous in the train frame. The lightning strike at the front of the train must occur first and the wavefronts from the two flashes do not reach the center of the train at the same time in either frame. In this way, the relativity of simultaneity is seen to be a consequence of the invariance of the speed of light.

*Description of tutorial sequence on the train paradox*

The original version of the train paradox used in tutorial differs slightly from the one described above. Students are told that two sparks occur at either end of a train that moves with relativistic speed relative to the ground. The sparks leave char marks on the ground and on the train.[14] The ground-based observer, Alan, who is at rest midway between the marks on the ground, receives the wavefronts from the sparks at the same time. (See Fig. 2(a).)

*Analysis in ground frame:* Students are asked to draw a diagram for the ground frame that shows the wavefront of light from each spark shortly after the sparks occur. They are guided to recognize that the wavefronts from both sparks are spheres centered on the char marks on the ground (since the propagation of light is isotropic) and that they are the same size in the ground frame (since they reach Alan at the same time). Students are then told that an observer, Beth, is standing at the center of the train. They are asked whether, in Alan's reference frame, Beth receives the wavefront from the front spark (wavefront F) before, after, or at the same time as the wavefront from the rear spark (wavefront R). Most students recognize that Beth receives wavefront F before wavefront R since in Alan's frame she is moving toward the center of the front wavefront. A correct diagram for the situation in Alan's frame is shown in Fig. 2(b).

*Analysis in train frame:* The students are next asked to determine the order of the events in the train frame. A correct answer involves recognizing that in the train frame, as in the ground frame, Beth receives wavefront F before wavefront R. In the train frame, the train is at rest and thus the wavefronts from the sparks are spheres centered on the char marks at the ends of the train. Since wavefront F reaches Beth's location first in her frame, and in her frame she is equidistant from the event locations, the front spark must occur first in her frame.

## B. Identifying and addressing student difficulties related to causality in the context of the train paradox

We had not anticipated the extent to which the transition from the ground frame to the train frame would be challenging for students. Our observations of students in the classroom, however, indicate that the transition is very difficult for students when they are required to construct the resolution of the paradox themselves. Most students answer (correctly) that, in Alan's reference frame, the wavefronts from the two sparks reach Beth at different times. They then answer (incorrectly) that, in Beth's reference frame, the wavefronts reach her at the same time. This is the essence of the paradox discussed above. However, very few students recognize an inconsistency in these two answers. Most students simply move on to subsequent activities in the tutorial. They do not see the logical necessity of the relativity of simultaneity and thus do not confront their belief that simultaneity is absolute.

The answers given by the students indicate a failure to recognize that two events that occur at a single location (*e.g.,* the receptions of two flashes by Beth) must have the same time order in all reference frames. The preservation of the order of the receptions of the wavefronts in the two frames is implicit in the resolution of the train paradox given above. The requirement that the two flashes reach Beth in the same order in all reference frames is a consequence of causality. [If the time (**d***t*) between two events is sufficient for a light signal to propagate between their locations (**d***x*), i.e., $c^2 > dx^2/dt^2$ or $ds^2 = c^2 dt^2 - dx^2 > 0$, then those events have a time-like separation and a possible causal relationship. Therefore, the time order in which they occur must be the same in all frames. If the time order could be reversed or made zero then the 'result' could precede the 'cause.'] Since the two events corresponding to the reception of the wavefronts by Beth have a time-like separation in the ground frame they occur in the same order in all frames and cannot be simultaneous in any frame.

We decided to modify the tutorial to help students recognize the 'paradox' in the train paradox. The approach we took was to shift the focus from the time order of two events (the reception of each wavefront) to whether or not a single event occurs.[15]

### 1. Eliciting difficulties with causality

In the modified tutorial, students are told that Beth has a tape player that operates as follows. When wavefront F reaches the tape player, it starts to play music at top volume. When wavefront R reaches it, the tape player is silenced. If both wavefronts reach the tape player at the same instant, it remains silent. Students are asked whether the tape player plays (i) in Alan's frame and (ii) in Beth's frame. The analysis in Alan's frame (described above) shows that Beth receives wavefront F before wavefront R, and thus the tape player plays. Causality requires that the tape player plays in the train frame as well.

The tape player exercise leads students to recognize that different answers about the order in which Beth receives the wavefronts in the two frames results in different answers about whether or not a particular event occurs. We found that the exercise with the tape player helps students confront the 'paradox' in the train paradox. However, most students still have difficulty in resolving the situation on their own. Some specific difficulties elicited by the modified tutorial are discussed below.

- *Failure to recognize that events that occur in one frame occur in all frames*

The fact that the tape player plays in all frames is not immediately obvious to students. Instead, many claim that the music plays in the ground frame but not in the train frame. For most students, belief in absolute simultaneity seems to be sufficiently strong that they fail to consider the relativity of simultaneity in resolving the paradox.

Subsequent questions in the tutorial ask whether Beth will hear the music and whether Beth will later observe the tape to have advanced from its starting position.[16] Presented with such concrete physical applications of causality, students begin to recognize that they hold deeply incompatible beliefs about the physical world.[17,18] The following exchange between two advanced undergraduates and a physics graduate student was recorded in the classroom.[19]

S1: We just figured out that the tape player plays in Alan's frame.

S2: But it can't. In Beth's frame they [the wavefronts] hit her at the same time. So she won't hear it.

S3: But look down here, it's asking if she hears it and if the tape will have wound from its starting position. If the tape is going to play, that's it; it's going to play.

S2: But it can't play for Beth! She's in the middle. They hit her at the same time.

S1: But we just figured out that it plays!

The above exchange is typical of student interactions on this exercise. Students refute one another vigorously. Some reject the entire scenario as impossible, but most accept that the tape plays in Alan's frame but not in Beth's. They conclude, erroneously, that special relativity implies that events that occur in one frame do not necessarily occur in all frames. Few students recognize spontaneously that they can resolve the conflict by discarding absolute simultaneity. This is the case even after they have studied the relativity of simultaneity in class and have worked homework problems on this topic.

- *Tendency to treat different frames of reference as corresponding to different objective realities*

A common response by students is to invent an "alternative reality" in an attempt to reconcile conflicting ideas. The students in the following exchange brought in poorly-understood ideas from quantum mechanics to support the erroneous idea that the cassette tape player both plays and does not play.[20] ["I" indicates the instructor.]

S1: Wait, so Alan hears it and Beth doesn't? That's one awesome tape player.

S2: That's so cool.

I: But when you take the tape out, when you stop the train and you look at the tape, has it been wound or has it not been wound?

S1: This is what [the instructor was] telling us last week. That in some universe Sara was wearing purple and in another one she was wearing blue or something.

In a course for high school teachers, a student and the instructor came up with a modified scenario: If, in Beth's frame, she encounters the front wavefront first, then her hat flies out of the train and Alan picks it up and wears it. If she receives both wavefronts at the same time, her hat remains on her head. When the student was asked how many hats would be present during Alan's and Beth's reunion, he replied cautiously, "Two." The thoughtful acquiescence of the student's partner further confirmed for us the suspicion that students do not recognize the crucial choice to be made: allow events to occur in one frame and not in another (a violation of causality) or abandon absolute simultaneity. They act as if the former were the only possible option.

In interview situations, where there are no classmates with whom to discuss the intellectual conflict, many otherwise animated students respond to the tape player scenario with silence.[21] In contrast to other occasions during the interview, students tend not to articulate their thoughts, ask questions, or respond to statements by the interviewer. This nearly complete stillness can last for a long time (about thirty seconds).

The failure to consider the possibility that the two events are not simultaneous in Beth's frame (when signal travel time is taken into account) seemed to be equally prevalent among students who had or had not studied special relativity. Few students, after study of relativity, appear to have recognized the implications of the relativity of simultaneity, despite familiarity with the paradoxes intended to illustrate this idea.

## 2. Addressing difficulties with causality

Both in the classroom and in interviews, students appear to require time for reflection in order to resolve their difficulties. Students are often confounded when they leave their tutorial session, but come to accept the necessary conclusion once they have had time to repeat (several times) the multi-step reasoning in the tutorial and homework. The graduate students in the interviews eventually agree that the relativity of simultaneity is logically inevitable. Many have difficulty recalling their former reasoning. "I don't know what I was thinking," one stated. "The tape player has to play."

Once students accept the idea that the tape player plays in both frames, the remainder of the analysis follows quickly. Students illustrate their answer for Beth's frame with a diagram similar to that shown in Fig. 3, in which the wavefronts are centered on the ends of the train and the front wavefront is larger.

## 3. Commentary

We have observed with interest that difficulties with the consequences of causality rarely arise in traditional treatments of the relativity of simultaneity. We believe that this is so because many

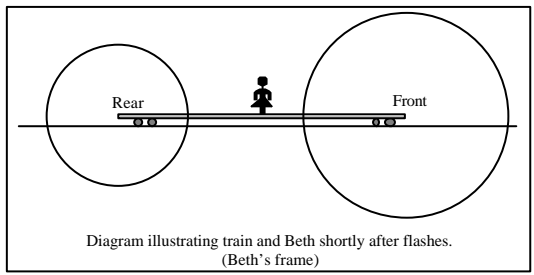

Diagram illustrating train and Beth shortly after flashes.
(Beth's frame)

Figure 3: Diagrams of train paradox for train-based observer. Example of correct diagram showing the wavefronts from the sparks that occurred at the ends of the train at the same time for a ground-based observer. The wavefronts are centered on the ends of the train, and the front spark occurs first.

students do not reach the level of sophistication required to consider them. For example, students who believe that simultaneity is a matter of signal perception readily accept that Beth records the events in a different time order than Alan does. Causality is irrelevant to their analysis.

### C. Cementing the relationship between relativity of simultaneity and reference frames in relative motion

In working through the exercises described above, many students start to change their understanding of simultaneity in a profound way. They need practice, however, in applying, extending, and generalizing the ideas to other contexts.

#### 1. Addressing the belief that every observer constitutes a different reference frame

We have found that it is crucial to have students reexamine their earlier conclusions about the meaning of reference frame in light of their new understanding of simultaneity. The tutorial describes an additional observer, Becky, at rest on the rear of the train and asks whether, in Becky's frame, the front spark occurs before, after, or at the same time as the rear spark. Students recognize that even though Becky sees wavefront R first, wavefront F is created first in her frame as it is in Beth's.

#### 2. Applying the relativity of simultaneity in new contexts

The ideas developed in the *Simultaneity* tutorial are counterintuitive. The tutorial helps students deepen their understanding by applying these concepts in a variety of other situations.

*Relativity of simultaneity as related to Lorentz contraction.* The *Simultaneity* tutorial typically comes after lecture instruction on Lorentz contraction. We have found that students often have little difficulty believing that the length of an object is greatest in its rest frame (although we have substantial evidence that students apply length contraction indiscriminately).[2] One tutorial exercise uses length contraction to reinforce the relativity of simultaneity. Students analyze a classic paradox in which two rods pass and are found to have the same length in the frame of one of the rods. They apply length contraction to show that the rods have different lengths in the frame of the other rod. They are led to recognize that the two events corresponding to the passing of the two ends are simultaneous in the frame of the first rod, but not in the frame of the second rod.

*Relativity of simultaneity as the resolution of another classic paradox.* In the homework for the *Simultaneity* tutorial, students consider a variation of a classic paradox. An object with a rest length greater than that of a container moves past the container at relativistic speed and seems to fit within the container. Students analyze the situation and show that the physical outcomes are consistent in the reference frames of both objects. A correct analysis requires application of the relativity of simultaneity. (The exercise also illustrates for students the impossibility of perfect rigidity in special relativity.)

### VI. ASSESSING STUDENT UNDERSTANDING OF SIMULTANEITY

Ongoing assessment of student learning plays a critical role in the development of curriculum by the Physics Education Group.

Below, we discuss results from three questions that have been administered before and after tutorial instruction to assess student understanding of time, reference frames, and simultaneity.[22] On each question, student performance in different courses at the same level was similar. Therefore, in the following discussion, the results have been combined. No student saw the same version of any question as both a pretest and post-test.

### A. Assessing student understanding of reference frames: *Seismologist* question

As discussed previously, students often fail to treat a reference frame as a set of observers who agree on the time order of events. One question that we have used in our investigation examines whether or not students distinguish the time order of two distant events from the time order in which an observer receives signals from the events. Many versions have been given. They are collectively entitled the *Seismologist* question. One version is discussed below.

#### *1. Description of the question*

In the *Seismologist* question, two volcanoes, Mt. Rainier and Mt. Hood, suddenly erupt and a seismologist at rest midway between them sees the eruptions at the same instant. A second observer (the "assistant") is at rest relative to the ground at the base of Mt. Rainier. Students are asked whether Mt. Rainier erupts before, after, or at the same instant as Mt. Hood in the reference frame of the assistant.

To answer correctly, students must be able to apply the definition of simultaneity and understand the role of a reference frame in establishing a common time coordinate for observers at rest relative to one another. The seismologist is equidistant from the mountains, so the signal travel times are the same and thus occurred at the same time. Since both observers are in the same reference frame, they obtain the same answer for the order of the eruptions.

---

Two volcanoes, Mt. Rainier and Mt. Hood, are 300 km apart in their rest frame. Each erupts suddenly in a burst of light. A seismologist at rest in a laboratory midway between the volcanoes receives the light signals from the volcanoes at the same time. The seismologist's assistant is at rest in a lab at the base of Mt. Rainier.[*]

Define Event 1 to be "Mt. Rainier erupts," and Event 2 to be "Mt. Hood erupts."

All observers are *intelligent* observers, *i.e.,* they correct for signal travel time to determine the time of events in their reference frame. Each observer has synchronized clocks with all other observers in his or her reference frame.

For the intelligent observer at the base of Mt. Rainier, does Event 1 occur *before, after,* or *at the same time as* Event 2? Explain.

[*]*In this problem, all events and motions occur along a single line in space. Non-inertial effects on the surface of the Earth may be neglected.*

Figure 4: The *Seismologist* question.

#### *2. Administration of the question*

We have given the *Seismologist* question to undergraduate students before and after traditional instruction, as well as after the pair of tutorials *Events and reference frames* and *Simultaneity*. The question has also been given to advanced undergraduates and graduate students during in-depth individual demonstration interviews and to physics graduate students as part of a question on a physics qualifying examination at the UW.

#### *3. Student performance*

Without tutorial instruction, relatively few undergraduates (between 20%-30% at the introductory level, and about 40% at the advanced level) answered correctly about the time order of events in the frame of the assistant. (See the first four columns of Table I.) Student responses were similar before and after lecture instruction. The physics graduate students also had difficulty with this question. Only about two-thirds answered correctly on both the interviews and the qualifying examination. (See the fifth and sixth columns of Table I.)

Table I: Student performance on the *Seismologist* question: (a) without tutorial instruction (before and after traditional instruction) and (b) after tutorial instruction. [Percentages have been rounded to the nearest 5%.]

| | (a) Without tutorial instruction | | | | | | (b) With tutorial instruction | |
|---|---|---|---|---|---|---|---|---|
| | Before instruction | | After traditional instruction | | | | | |
| | Introductory students | Advanced undergraduates | Introductory students | Advanced undergraduates | Graduate students (on qualifying examination) | Advanced undergraduates and graduate students (in interviews) | Introductory students | Advanced undergraduates |
| | ($N = 88$) | ($N = 48$) | ($N = 79$) | ($N = 63$) | ($N = 23$) | ($N = 17$) | ($N = 197$) | ($N = 98$) |
| | % (N) | % (N) | % (N) | % (N) | % (N) | % (N) | % (N) | % (N) |
| Correct answer (simultaneous eruptions) regardless of reasoning | 20% (19) | 40% (20) | 30% (25) | 40% (24) | 65% (15) | 60% (10) | 85% (167) | 85% (82) |
| Rainier erupts first | 65% (57) | 55% (26) | 60% (49) | 50% (33) | 35% (8) | 40% (7) | 15% (28) | 15% (14) |
| Other (*e.g.*, Hood erupts first, student stated not enough information given) | 15% (12) | 5% (2) | 5% (5) | 10% (6) | 0% (0) | 0% (0) | 0 (0) | <5% (2) |

The most common incorrect answer is that the events are not simultaneous for the assistant. This response is consistent with a belief that the time order of events depends on the order in which an observer receives signals from the events. In effect, the students treat observers at rest relative to one another as being in different reference frames.

After students have completed the two tutorials, performance on this question is very good. About 85% of the introductory and advanced undergraduate students answered correctly. (See the last two columns of Table I.) This is better than the performance of the graduate students on the qualifying examination. The undergraduates who responded incorrectly after tutorial instruction (~15%) gave answers similar to those by students before tutorial instruction.

### B. Assessing student understanding of the relativity of simultaneity: *Spacecraft* question

Some of the questions used to assess the effectiveness of the tutorials allow us to probe the extent to which students can apply the relativity of simultaneity. One such question, entitled the *Spacecraft* question, is discussed below.

#### 1. Description of the question

The *Spacecraft* question involves two volcanoes, Mt. Rainier and Mt. Hood, which erupt simultaneously according to an observer at rest on the ground midway between them. The question states that a spacecraft is flying at relativistic velocity from Mt. Rainier to Mt. Hood and is over Mt. Rainier at the instant it erupts. The eruption events are explicitly labeled Event 1 (Mt. Rainier erupts) and Event 2 (Mt. Hood erupts). Students are asked whether, in the reference frame of the spacecraft, Event 1 occurs before, after, or at the same time as Event 2.

A correct answer can be obtained through the use of qualitative or quantitative reasoning or from a spacetime diagram. The following is an example of a qualitative argument that we accept as correct. In the spacecraft frame, the locations at which the eruptions occur are stationary. We can imagine these as the centers of wavefronts of light from the eruptions. According to an observer in the spacecraft, the ground-based observer is moving away from the center of

> Mt Rainier and Mt. Hood, which are 300 km apart in their rest frame, suddenly erupt at the same time in the reference frame of a seismologist at rest in a laboratory midway between the volcanoes. A fast spacecraft flying with constant speed $v = 0.8c$ from Rainier toward Hood is directly over Mt. Rainier when it erupts.
>
> Let Event 1 be "Mt Rainier erupts," and Event 2 be "Mt. Hood erupts."
>
> In the reference frame of the spacecraft, does Event 1 occur *before, after,* or *at the same time as* Event 2? Explain your reasoning.

Figure 5: The *Spacecraft* question.

> Two harmless explosions occur at the ends of a landing strip whose proper length is 3000 m. In the reference frame of the landing strip engineer (at rest on the strip), the first explosion occurs at the left end of the strip, and the second explosion occurs at the right end of the strip at a time $c\Delta t = 1200$ m later.
>
> Is there a reference frame in which the two explosions occur at the same instant? If so, determine the magnitude and direction of the velocity of this frame relative to the landing strip. If not, explain why not.

Figure 6: The *Explosions* question.

the flash from Mt. Hood and toward the center of the flash from Mt. Rainier. Thus, in the spacecraft frame, the ground-based observer is closer to the center of the signal from Mt. Rainier at the instant that observer receives both signals. In the spacecraft frame therefore, Mt. Hood erupted first since its signal travels farther in order to reach the ground-based observer at the same time as the signal from Mt. Rainier. A correct answer can also be obtained using the Lorentz transformation for time.[23]

### 2. Administration of the question

We have given versions of the *Spacecraft* question to undergraduate students after traditional instruction and after traditional and tutorial instruction on the relativity of simultaneity. The question has also been given to advanced undergraduates and graduate students during in-depth individual demonstration interviews and to physics graduate students on the physics qualifying examination.

### 3. Student performance

Student performance on the *Spacecraft* question before tutorial instruction is summarized in the first six columns of Table II. Performance at all levels is poor, both before and after traditional instruction. Fewer than 30% of the students in each population have given a correct response (with or without correct reasoning). Many students responded that Mt. Rainier erupts first for the spacecraft observer. They reason that the observer is closer to Mt. Rainier and would thus see it erupt first. Other students recognized that signal travel time should be taken into account, but often claimed that after doing so the events would be simultaneous in the spacecraft reference frame.

Both introductory and advanced students seem to benefit from working through the tutorials. About half of each group answered correctly on the *Spacecraft* question when it was given after tutorial instruction. The tendency to reason on the basis of absolute simultaneity or to reason solely on the basis of signal reception time decreased for both populations. Both populations did substantially better than graduate students who had not had tutorial instruction.[24,25] Thus, the tutorial sequence seems to be successful in helping students develop a better understanding of simultaneity and reference frames.

### C. Assessing student ability to solve quantitative problems requiring use of relativity of simultaneity

Some of the assessment questions we have used are quantitative. Below, we discuss student performance on a question entitled the *Explosions* question that can be

The challenge of changing deeply-held student beliefs… May 2002
Scherr, et al. 12

Table II: Student performance on the *Spacecraft* question: (a) before and after traditional instruction and (b) after tutorial instruction. [Percentages have been rounded to the nearest 5%.]

| | (a) Without tutorial instruction | | | | | | (b) With tutorial instruction | |
|---|---|---|---|---|---|---|---|---|
| | Before instruction | | After traditional instruction on relativity of simultaneity | | | | | |
| | Introductory students | Advanced undergraduates | Introductory students | Advanced undergraduates | Graduate students (on qualifying examination) | Advanced undergraduates and graduate students (in interviews) | Introductory students | Advanced undergraduates |
| | ($N = 67$) | ($N = 20$) | ($N = 73$) | ($N = 93$) | ($N = 23$) | ($N = 11$) | ($N = 173$) | ($N = 70$) |
| | % (N) | % (N) | % (N) | % (N) | % (N) | % (N) | % (N) | % (N) |
| Correct answer: Hood erupts first (with correct reasoning *or* incomplete reasoning *) | 5% (3) | 15% (3) | 10% (8) | 25% (24) | 30% (7) | 25% (3) | 50% (89) | 55% (38) |
| Simultaneous eruptions (reasoning consistent with being based on absolute simultaneity) | 20% (12) | 25% (5) | 5% (5) | 20% (19) | 10% (2) | 0 (0) | <5% (2) | 10% (8) |
| Rainier erupts first (reasoning consistent with being based on the times at which signals are received by the observer) | 70% (46) | 45% (9) | 75% (55) | 40% (39) | 60% (14) | 55% (6) | 40% (70) | 35% (24) |
| Other (*e.g.*, student stated not enough information given) | 10% (6) | 15% (3) | 5% (5) | 10% (11) | 0 (0) | 20% (2) | 5% (12) | 0 (0) |

* Some students gave a correct answer with reasoning that was incomplete, but not incorrect. Although it was not possible to tell whether they were correct in their reasoning, in this article the responses are treated as correct.

solved through application of the Lorentz transformations.

#### 1. Description of the question

In the *Explosions* question, an explosion occurs at each end of a landing strip with a proper length of 3000 m. In the frame of an engineer at rest on the strip, the explosion at the right end occurs a time *dt* after the explosion on the left end (where $cdt = 1200$ m). Students are asked whether there is a frame in which the explosions are simultaneous, and if so, to determine the velocity of that frame relative to the landing strip.

A correct answer can be found through use of the Lorentz transformations. The spatial separation between the explosions *(dx)* is 3000 m and the time separation *(dt)* corresponds to 1200 m. Thus, the time duration between the explosions is zero in a frame that moves from left to right with speed $0.4c$.

#### 2. Administration of the question

The *Explosions* question has been given on examinations after standard instruction to introductory students ($N = 128$) and advanced undergraduates ($N = 31$). It has also been used in interviews with undergraduate and graduate students ($N = 17$) after standard instruction. The question has been administered after tutorial instruction on examinations to introductory students ($N = 84$) and advanced undergraduates ($N = 25$).

#### 3. Student performance

After traditional instruction, about 45% of the introductory students and about 30% of the advanced undergraduates answered the *Explosions* question correctly. The mathematical nature of the question made student errors difficult to categorize. However, in many cases, conceptual difficulties seemed to prevent students from answering correctly. For example, some students claimed that the location of the moving observer would determine the order of events for that observer.

After working through the pair of tutorials described, about 60% of introductory students and 70% of advanced undergraduates answered correctly. This performance is comparable to that of graduate students (after traditional instruction) in an interview version of the task, on which 7 of 12 (about 60%) answered correctly.

The results suggest that the small investment of time (~2 hours) required by the tutorials can improve student ability to solve quantitative problems, although the small number of students in this study allows only for preliminary conclusions. Time spent in class on the tutorials on special relativity typically means that students spend less time in solving standard text-book problems. The findings suggest that addressing student conceptual difficulties can improve student performance on quantitative questions. This result is consistent with those obtained by our group in other topic areas.[26]

### D. Commentary

It should be noted that all the classes in which the three questions were administered after traditional instruction had included lectures in which a reference frame was defined through a system of intelligent observers and/or a set of clocks and metersticks. The students had seen a similar discussion in their textbooks. The previous paper describes how we repeatedly modified the questions to try to make clear to students that they should treat all observers as intelligent observers who take into account the signal travel time.[1] During interviews, the interviewer attempted to correct misinterpretations. Students at all levels held strongly to their ideas of time and reference frames. This observation guided the development of the tutorials on special relativity. The post-test results corroborate our finding that the specific student difficulties are very persistent and resistant to change.

## VII. CONCLUSION

The results of the investigation reported in this and a previous article indicate that many students who study special relativity at the undergraduate to graduate levels fail to develop a functional understanding. Even in advanced courses, students often do not recognize the implications of special relativity for our interpretation of the physical world. As in other advanced topics, we found that many student difficulties with this material could be traced to a lack of understanding of more basic, underlying concepts.[27]

In the two articles, we have shown how, through research, we were able to identify some conceptual hurdles that hinder students from applying basic kinematical concepts to the complex situations encountered in special relativity. After standard instruction many students lack operational definitions for such fundamental ideas as time of an event, simultaneity, and reference frame – concepts that should be familiar to them from Galilean relativity. We have illustrated how the results from research guided us in designing two tutorials (part of a larger set on relativity) that help students develop a sound understanding of these basic ideas. Students who had worked through these instructional materials improved significantly in their ability to recognize and resolve some of the classic paradoxes of special relativity.

In the traditional approach, paradoxes are often used as elicitation activities or motivational tools. However, a strategy in which the instructor elicits and exposes student beliefs to generate cognitive conflict and then resolves the paradox is inadequate. Our experience indicates that confrontation and resolution must be carried out by the students, not by the instructor, if meaningful learning is to take place. This strategy is especially crucial when the ideas are as strongly counterintuitive as in special relativity.

**ACKNOWLEDGMENTS**

The investigation described in this paper has been a collaborative effort by many members of the Physics Education Group, present and past. Bradley S. Ambrose and Andrew Boudreaux played significant roles in the initial stages of the research. We are especially grateful for the intellectual contributions of Lillian C. McDermott. Special thanks are due to Paula R.L. Heron and Mark N. McDermott. Also deeply appreciated is the ongoing cooperation of the colleagues in whose physics classes the instructional materials have been used, especially James Bardeen and E. Norval Fortson. The authors gratefully acknowledge the support of the National Science Foundation through Grants DUE 9354501 and DUE 9727648.

important to separate, as Einstein did in his 1905 paper, *local* from *distant* simultaneity. (See the article in Ref. 5.) Two distant events are *defined* to be simultaneous if their time coordinates in a specific reference frame are identical. This definition presupposes a *definition* for the time coordinate of a single event in a reference frame. This is most naturally *defined* as the reading on a clock located at the event's position "at the instant at which the event occurs." The concept of local simultaneity (the identification of the time of the event in question with the time that a local clock reads "at that instant") is assumed, therefore, to be a notion that does not require definition. Furthermore, to establish a particular clock reading for an event as the time coordinate of the event throughout a whole reference frame, a measurement procedure for how time may be "spread over space" needs to be specified. Our approach is consistent with those described in other texts. See, for example, P.W. Bridgman, *A sophisticate's primer of relativity* (Wesleyan University Press, Middletown, CT, 1962) and A.B. Arons, *A guide to introductory physics teaching* (Wiley, New York, NY, 1990).

[13] For a discussion of various instructional strategies by the Physics Education Group, including *elicit, confront, and resolve*, see L.C. McDermott, Oersted Medal Lecture: "Physics Education Research – The Key to Student Learning," Am. J. Phys. **69**, 1127–1137 (2001) and L.C. McDermott, Millikan Award Lecture: "What we teach and what is learned– Closing the gap," Am. J. Phys. **59**, 301–315 (1991).

[14] We are indebted to E.F. Taylor for numerous discussions that led us to incorporate the char marks into our instructional approach.

[15] We did not use an approach based on the invariant interval $(ds^2)$ since in most courses on special relativity time-like, space-like, and light-like intervals are discussed after the relativity of simultaneity.

[16] The fact that the music will be Doppler shifted is something that is not germane to the logical structure of the tutorial. Few students raise the issue.

[17] For a theoretical discussion of the circumstances under which encounters with new ideas produce dissatisfaction with an existing conception, see the last article in Ref. 3 and K.A. Strike and G.J. Posner, "A revisionist theory of conceptual change" in *Cognitive Psychology and educational theory and practice,* Duschl, Strike, and Hamilton (eds.), SUNY press, Albany, NY (1992).

[18] Cognitive disequilibrium and the approach toward equilibration is a major issue in developmental psychology. For examples of how children return to equilibrium through *assimilation, accommodation,* and *adaptation,* see J. Piaget, *The moral judgement of the child,* Free Press, New York, NY (1965); B. Rogoff, *Apprenticeship in thinking: Cognitive development in social context,* Oxford University Press, New York, NY (1990); P.H. Miller, *Theories of Developmental Psychology,* W.H. Freeman and Co., New York (1993); A.N. Perret-Clermont, *Social interaction and cognitive development in children,* Academic Press, New York, NY (1980).

[19] The conversation took place in a course for prospective high school science teachers. S1 and S3 are advanced undergraduate physics students; S2 is a first-year graduate student in physics. The course used an adaptation of the tutorial sequence that is being developed for *Physics by Inquiry,* a laboratory-based curriculum for the preparation of K-12 teachers. (L.C. McDermott and the Physics Education Group at the University of Washington, *Physics by Inquiry,* Vols. I and II, (Wiley, New York, NY, 1996.))

[20] This conversation was recorded in a modern physics course in a California high school that served as a pilot site for the *Events and reference frames* and *Simultaneity* tutorials.

[21] The interviews are discussed in Ref. 1. In addition to serving as a setting for probing student ideas about simultaneity, the interviews often helped us in identifying contexts and lines of questioning that might be effective as instructional strategies. These were eventually incorporated in the *Events and reference frames* and *Simultaneity* tutorials.

[22] We have found each question to be useful in eliciting specific student difficulties. For a detailed discussion about the development of the questions, see Ref. 1.

[23] An analysis based on the Lorentz transformations is given in Ref. 1.

[24] The graduate student data is for the *explicit* version of the Spacecraft question, which is similar but not identical to the tutorial post-test (the *location-specific* version). See Ref. 1 for a detailed discussion of each version of the Spacecraft question.

[25] For other examples in which undergraduate students perform, after tutorial instruction, as well as or better than graduate students without tutorial instruction, see the last article in Ref. 9. See also, S. Vokos, P.S. Shaffer, B.S. Ambrose, and L.C. McDermott, "Student understanding of the wave nature of matter: Diffraction and interference of particles," Phys. Educ. Res., Am. J. Phys. Suppl. **68**, S42-S51 (July 2000); B.S. Ambrose, P.S. Shaffer, R.N. Steinberg, and L.C. McDermott, "An investigation of student understanding of single-slit diffraction and double-slit interference," Am. J. Phys. **67**, 146-155 (1999); K. Wosilait, P.R.L. Heron, P.S. Shaffer, and L.C. McDermott, "Development of a research-based tutorial on light and shadow," *ibid.* **66**, 906-913 (1999).

[26] For an example in another areas, see, K. Wosilait, P.R.L. Heron, P.S. Shaffer, and L.C. McDermott, "Addressing student difficulties in applying a wave model to the interference and diffraction of light," Phys. Educ. Res., Am. J. Phys. Suppl. **67**, S5 – S15 (July 1999) and the last article in Ref. 9.

[27] For other research by our group consistent with this statement, see, for example, the third article in Ref. 25.